\documentstyle[12pt]{article}
\input epsf

\begin{document}
\date{\today}
\def\be{\begin{equation}}
\def\ee{\end{equation}}
\def\bea{\begin{eqnarray}}
\def\eea{\end{eqnarray}}
\def\N{{\mathcal{N}}}
\def\l{\lambda}
\def\e{\epsilon}
\def\g{\gamma}
\def\hg{\hat{\gamma}}
\def\he{\hat{e}}
\def\hs{\hat{s}}
\def\slashed#1{{\ooalign{\hfil\hfil/\hfil\cr $#1$}}}
\def\w{\omega}
\def\hw{\hat{\omega}}
\def\na{\nabla}
\def\hna{\hat{\nabla}}
\def\tK{\tilde{K}}
\def\d{\delta}
\def\G{\Gamma}
\def\C{\relax{\rm I\kern-.5em C}} 
\def\P{\relax{\rm I\kern-.5em P}} 
\def\tr{{\rm tr}}
\def\bP{{\bf {P}}}
\def\t{{\triangle}}


\newpage
\bigskip
\hskip 3.7in\vbox{\baselineskip12pt
\hbox{NSF-ITP-01-16}}

\bigskip\bigskip

\centerline{\large \bf  Comments on supergravity dual of pure $\N=1$}
\centerline{\large \bf Super Yang Mills theory with unbroken chiral symmetry}

\bigskip\bigskip

\centerline{{\bf
Alex Buchel\footnote{buchel@itp.ucsb.edu},
Andrew Frey\footnote{frey@physics.ucsb.edu}}}

\bigskip
\centerline{$^{1}$Institute for Theoretical Physics}
\centerline{University of California}
\centerline{Santa Barbara, CA\ \ 93106-4030, U.S.A.}

\bigskip
\centerline{$^2$Department of Physics}
\centerline{University of California}
\centerline{Santa Barbara, CA\ \ 93106, U.S.A.}
\bigskip

\begin{abstract}
\baselineskip=16pt
Maldacena and Nunez [hep-th/0008001] identified a gravity solution
describing pure $\N=1$ Yang-Mills (YM) in the IR. Their (smooth)
supergravity solution exhibits confinement and the $U(1)_R$ chiral
symmetry breaking of the dual YM theory, while the singular solution
corresponds to the gauge theory phase with unbroken $U(1)_R$ chiral
symmetry.  In this paper we discuss supersymmetric type IIB
compactifications on resolved conifolds with torsion.  We rederive
singular background of [hep-th/0008001] directly from the
supersymmetry conditions. This solution is the relevant starting point
to study non-BPS backgrounds dual to the high temperature phase of
pure YM. We construct the simplest black hole solution in this
background. We argue that it has a regular Schwarzschild horizon and
provides a resolution of the IR singularity of the chirally symmetric
extremal solution as suggested in [hep-th/0011146].

\end{abstract}
\newpage
\setcounter{footnote}{0}

\baselineskip=17pt

\section{Introduction}
The AdS/CFT duality of Maldacena \cite{juan} is a very useful tool 
in study of nonperturbative dynamics of four dimensional gauge theories.
The main idea of the approach is to use a dual gravitational description 
of the gauge theory living on a large stack of $N$ coincident 
D3-branes in string theory. When D3 branes are placed in a smooth 
type IIB background, the string theory in the near 
horizon geometry of the stack is dual to 
$\N=4$ supersymmetric YM theory \cite{juan,edholo,GPK}.

Several approaches are used to construct gravitational backgrounds 
dual to  gauge theories with reduced supersymmetry (and thus more interesting 
IR dynamics) \cite{agmoo}. In particular, $\N=1$ gauge theory can 
be obtained by mass deformation of parent $\N=4$ gauge theory, 
placing large number of D3 branes on appropriate conical singularity,
or, as suggested by Maldacena and Nunez \cite{mn0008}, in the IR 
of a little string theory realized by wrapping NS 5 branes of type 
IIB string theory on a 2-cycle, and appropriately twisting the 
normal bundle to preserve $1/4$ of original supersymmetries. 
Typically, in gravitational dual of nonconformal gauge theories with 
reduced supersymmetry one encounters naked singularities 
in the IR region. Over the last year we learned that these 
naked singularities potentially  signal to an interesting physical 
phenomenon in the IR dynamics of gauge theories. The nontrivial 
gauge theory IR physics often has a gravity (or string theory)
dual that resolves the naked singularity. Alternatively, understanding the 
resolution of naked singularities in gravitational backgrounds
\footnote{Not all naked singularities are physical and can be resolved.}
could teach us about nonperturbative effects in the gauge theory.

For example, from the 
chiral symmetry breaking in the IR of the gauge theory on the 
world volume of regular and fractional branes on the conifold, 
Klebanov and Strassler \cite{ks0007} argued that the 
naked singularity of the Klebanov-Tseytlin (KT) geometry 
\cite{KT} should be resolved via the deformation of the conifold.  
``The flow of information'' in the opposite direction,
i.e. from the string theory to the field theory, was proposed      
in \cite{b0011}. It was suggested there that naked singularity in 
the KT geometry could be alternatively resolved by placing 
sufficiently large black hole in the background, so that its horizon 
would cloak the naked singularity. ``Sufficiently large'' means that 
the Hawking temperature of the black hole horizon should be larger  
than the critical temperature of the chiral symmetry breaking 
phase transition. Here, the hope is that, by studying the black hole 
of the critical radius one would learn about the gauge theory 
phase transition. The black hole solution proposed in \cite{b0011}
fails to realize this scenario. As shown in \cite{new}, 
the horizon of the non-extremal 
solution presented  in \cite{b0011} does not cloak the singularity, but 
rather coincides with it. This type of singular horizon is deemed 
unacceptable in studies of black hole metrics. In \cite{new}
a system of second order equations is derived whose solutions 
may describe non-extremal generalizations of the 
KT background with regular horizons. Recently constructed 
smooth solutions to this system \cite{shkt} in a perturbation 
theory valid for large Hawking temperature of the horizon 
show that this is indeed possible. 
These solutions appear to support the suggestion of \cite{b0011}
that a regular horizon of the non-extremal generalization of 
the KT geometry appears only at some finite Hawking temperature.  

Gravitational backgrounds which regular Schwarzschild horizon 
exists only above some critical non-extremality are unusual and, to our 
knowledge,  new  from the supergravity point of view. 
Analysis of \cite{b0011,new,shkt}  suggests that they should 
nonetheless be quite generic for backgrounds dual (in  Maldacena sense) 
to gauge theories which undergo finite temperature symmetry breaking 
phase transition. It is thus of interest to look for additional examples of 
this phenomenon. An obvious choice  is the 
supergravity solution constructed 
by Chamseddine and Volkov in \cite{chv1,chv2} and 
interpreted by Maldacena and Nunez  \cite{mn0008} as a 
gravity dual of the pure 
$\N=1$ supersymmetric Yang Mills theory in the IR. 
In what follows we refer to this supergravity background 
as CV-MN.
The smooth solution of  \cite{chv1,chv2,mn0008} has the same  IR 
behavior as that of the cascading gauge theory of \cite{ks0007}, thus 
corresponding to the phase of the gauge theory with broken 
chiral symmetry at zero temperature. The singular CV-MN solution,
like the KT geometry, describes the phase of the gauge theory 
with the unbroken symmetry. The crucial difference between the two 
models is that KT-KS  model has an effective four dimensional 
description in the UV as well \cite{kr}, while the  
CV-MN model is regularized in the UV by the little string theory. 
As a first step towards 
understanding the thermodynamics and the phase transition in the 
CV-MN model we construct  a non-BPS generalization of the 
chirally symmetric (singular) CV-MN background. We argue 
that presented solutions have a regular horizon that exists 
above some critical value of non-extremality, in agreement 
with the dual gauge theory where the chiral symmetry is 
restored at finite temperature. 

As a separate issue, we will also extend a no-go theorem for
supersymmetric compactifications with torsion \cite{andy86,compact}.
The work of \cite{compact} studied compactifications of type IIB 
supergravity of the form given by \cite{andy86} to 4 Minkowski
dimensions.  They showed that for 6 compact internal dimensions, the
only solutions with globally defined dilaton are Calabi-Yau three-folds
with vanishing torsion.  This no-go theorem can be avoided on noncompact
internal manifolds, as the CV-MN solution shows.  Our results extend the
no-go theorem of \cite{compact} to noncompact internal manifolds with
the complex structure of the resolved conifold, in that all BPS solutions
will have a naked singularity.

This paper is organized as follows. 
In the next section following the work \cite{andy86}, we 
review the general construction of supersymmetric 
vacua of type IIB supergravity with torsion. In section 
3 we study supersymmetric compactifications of type IIB
supergravity on resolved conifolds (which is relevant to the 
unbroken chiral symmetry phase of the CV-MN model). We show 
that all SUSY preserving vacua have naked singularity.  
In section 4 we rederive a simple singular solution of
\cite{chv1,chv2,mn0008}. In section 
5 we  discuss non-extremal generalizations
of the background. We conclude with discussion in 
section 6.

\section{Supersymmetric type IIB compactifications with torsion}

Conditions for spacetime supersymmetry of the 
heterotic superstring on manifolds with torsion were found in 
\cite{andy86}. In this section we consider corresponding conditions 
for type IIB compactifications to four dimensions on manifolds with 
torsion. As we set all the R-R fields to be zero, our analysis essentially 
repeat those of \cite{andy86}. Similar to the heterotic 
compactifications, we find that supersymmetric 
type IIB vacua with torsion have warped 
four dimensional space-time and nontrivial dilaton.

Type IIB equations of motion and supersymmetry variations 
have been found in \cite{schwarz83} which notation  we follow 
here\footnote{See Appendix  for our detailed notations 
and conventions.}. 
In particular, we use mostly negative signature for the metric. 
The massless bosonic fields of the type IIB superstring theory consist of
the complex dilaton field  $B$ that
parameterizes the $SL\left(2,\bf R\right)/ U\left(1\right)$ coset space, 
the metric tensor $g_{MN}$ and the antisymmetric complex 
2-tensor $A_{(2)}$, and the
four-form field $A_{(4)}$ with self-dual five-form field strength. 
Their fermionic superpartners are a complex Weyl
gravitino $\psi_{M}$ ($\hg^{11}\psi_{M}=-\psi_{M}$) and a complex Weyl
dilatino $\lambda$ ($\hg^{11}\lambda=\lambda$). The theory has 
$\N$=2 supersymmetry generated by two supercharges of the same chirality.

We would like to find bosonic backgrounds that preserve 
some supersymmetry. This will be the case provided the supersymmetry 
variation of the fermionic fields is zero 
\bea
\delta\l&=&i P_M\hg^M\e^{*}-{i\over 24 }G_{MNP}\hg^{MNP}\e\,,
\cr
\cr
\delta\psi_M&=&\hat{D}_M \epsilon + 
     {i \over 480} F_{P_1 \cdots P_5} \hg^{P_1 \cdots P_5}
     \hg_M \epsilon + {1 \over 96} (\hg_M{}^{NPQ} G_{NPQ}\cr
&&\qquad\qquad\qquad -      9 \hg^{NP} G_{MNP}) \epsilon^*\,,
\label{susy}
\eea 
where 
\bea
   &&F_{(5)} = dA_{(4)} - {1 \over 8} {\rm Im} A_{(2)} \wedge F^*_{(3)}\,, 
\qquad
   F_{(3)} = dA_{(2)}\,,  \cr\cr
   &&G_{(3)} = {F_{(3)} - B F^*_{(3)} \over \sqrt{1-|B|^2}}\,,  \qquad
   P_M = {\partial_M B \over 1-|B|^2}\,,\cr
\cr
&&\hg^{11}\e=-\e\,. 
\eea
The covariant derivative $\hat{D}_M$ contains $U(1)$  connection
$Q_M={\rm Im}(B\partial_M B^*)/(1-|B|^2)$
\be
\hat{D}_M\e=\left(\hna_M-{1\over 2}i Q_M\right)\e=
\left(\partial_M-{1\over 4}\hw_{MNP}\hg^{NP}-{1\over 2 }i 
Q_M\right)\e\,.
\label{covder}
\ee
The spin connection is given by
\be
\hw_{MNP}=\he_M^r{}_[\partial_N\he_P{}_]{}_r+\he_P^r{}_[\partial_N\he_M{}_]{}_r
+\he_N^r{}_[\partial_M\he_P{}_]{}_r\,.
\label{spin}
\ee
The above supergravity potentials and the dilaton field differ from those 
conventionally used in D-brane physics. In the latter case we have 
the dilaton $\phi$, $B_{(2)}$ two-form from the NS-NS sector, and 
$C_{(n)}$ forms from the R-R sector with $n=0,2,4,6,8$. The dictionary 
between the two descriptions was presented in \cite{bpp}, which we adopt 
here. In particular we have 
\bea
&&C_{(0)}+ie^{-\phi}=i{1+B\over 1-B}\,,\cr
\cr
&&A_{(2)}=C_{(2)}+i B_{(2)}\,.
\label{dic}
\eea 
It is easy to see that in type IIB equations of motion we can consistently 
set all R-R potentials to zero. Thus we have 
\bea
&&G_{MNP}=i H_{MNP} e^{-\phi/2}\,,\qquad P_M=-{1\over 2}\partial_M\phi\,,\cr 
&&F_{P_1\cdots P_5}=0\,,\qquad Q_M=0\,,
\label{noRR}
\eea
where $H=d B_{(2)}$. We assume geometry to be a direct product of two
spaces $M_4\times K$
\be
d\hs_{10}^2=e^{2A(y)}\eta_{\mu\nu}dx^{\mu}dx^{\nu}-e^{2B(y)}
g_{mn}dy^mdy^n\,,
\label{metric10}
\ee
with warp factors depending on the coordinates of the six dimensional 
factor only. The warp factor $B(y)$ is not to be confused with the 
complex dilaton parameterizing  $SL\left(2,\bf R\right)/ U\left(1\right)$ 
coset space;  we will not use the latter in the following. 
Furthermore, we take both the dilaton and the NS-NS two form to 
depend only on the coordinates $y^m$ of $K$. 
Introducing 
\be
\e=e^{i\pi/4}\eta\,,\qquad \eta^*=\eta\,,
\ee
and manipulating with gamma matrices, 
supersymmetry variations (\ref{susy}) become
\bea
&&\delta\l=\left[-{1\over 2}\hat\slashed{\partial\phi}+{1\over 4}e^{-\phi/2}
\hat{\slashed{H}}\right]\eta\,,\cr
\cr
&&\delta\psi_M=\hat\nabla_M\eta-{e^{-\phi/2}\over 16}
\left(2\hat\slashed H\hg_M+\hg_M\hat\slashed H\right)\eta\,.
\label{ss}
\eea
Vanishing of the dilatino and gravitino variations (\ref{ss})
is identical to the supersymmetry preserving conditions 
of the heterotic compactification with torsion discussed in
\cite{andy86}, provided we set gauge fields to zero.
So we can simply adopt the latter results.
First of all, from the gravitino variation in 
four dimensions we find  
\be
A=-{1\over 4} \phi\,,
\label{a}
\ee 
that is, the $M_4$ factor of the background geometry is
flat Minkowski space in string frame.  The remaining SUSY
preserving conditions is  convenient to formulate in 
string frame, that is choosing 
\be
B=-{1\over 4}\phi\,.
\label{b}
\ee
Consider six dimensional manifold $\tK$, with the
metric 
\be
ds^6_{\tK}=g_{mn} dy^m dy^n\,.
\label{tk}
\ee
Unbroken supersymmetry in four dimensions implies 
$\tK$ to be hermitean, endowed  with a holomorphic (3,0) form
$\w$ \cite{andy86}:
\be
\bar{\partial}\w=0\,.
\label{dw}
\ee
Introducing global complex coordinates 
$z^i$ and $\bar{z^i}\equiv z^{\bar{i}}$ on $\tK$
and denoting global (anti-)holomorphic indices 
as $(\bar{a},\bar{b},\bar{c},\cdots)\ a,b,c,\cdots$,
the fundamental (1,1) form on $\tK$
\bea
J&=&i g_{a\bar{b}} dz^a\wedge dz^{\bar b}\,
\label{jf}
\eea
relates to $\w$ as 
\be
d^+J+i\left(\bar{\partial}-\partial\right) \ln ||\w||=0\,,
\label{finale}
\ee 
where the norm of $\w$ is given by 
\be
||\w||^2=\w_{a_1a_2a_3}\bar{\w}_{\bar{b}_1\bar{b}_2\bar{b}_3}
g^{a_1\bar{b}_1}g^{a_2\bar{b}_2}g^{a_3\bar{b}_3}\,.
\label{wnorm}
\ee
Finally, the torsion is 
\be
H=i\left(\bar{\partial}-\partial\right)J\,,
\label{h}
\ee
and the dilaton is given by 
\be
\phi=\phi_0-{1\over 2} \ln||\w||\,,
\label{wfin}
\ee
where we explicitly included constant $\phi_0$.

\section{Supersymmetric compactifications on 
resolved  conifolds with  torsion}

{\footnote{Results reported in this section we obtained in 
collaboration with Joe Polchinski.}
Supersymmetric compactifications on Hermitian manifolds with torsion 
are very restrictive. In fact, in \cite{compact} it was shown 
that there are no supersymmetric compactifications of this type on 
compact Hermitian manifolds $M^6$ with non-vanishing torsion and a globally 
defined dilaton. This no-go theorem could be avoided with non-compact 
$M^6$ \cite{mn0008}. It is thus of interest to study 
supersymmetry on other non-compact Hermitian manifolds with torsion. 

In this section we discuss supersymmetric compactifications 
of type IIB supergravity on resolved conifolds with torsion.  
We explicitly show that with the nonvanishing torsion, the resolved 
conifold geometry always has a naked singularity, also the dilaton 
can not be defined globally.  As shown in \cite{mn0008} supersymmetric 
backgrounds in this class 
are dual to the phase with unbroken chiral symmetry 
of pure $\N=1$ four dimensional Yang Mills theory at zero temperature. 
Since the 
chiral symmetry of the YM theory is broken in the IR, the 
appearance of a naked singularity is not  
surprising.

The general supersymmetry conditions for type IIB compactification
with torsion were reviewed in the previous section. The string frame 
metric is given by 
\be
ds^2_{str}=\eta_{\mu\nu}dx^{\mu}dx^{\nu}-ds_6^2\,, 
\label{stmetric}
\ee 
where the hermitian metric $ds_6^2$ on the six dimensional manifold 
$\tK$ (resolved conifold in our case) satisfies 
(\ref{finale}) once the holomorphic 
$(3,0)$ form $\w$ is specified. The SUSY preserving 
torsion is then given by (\ref{h}), and the dilaton is determined 
from (\ref{wfin}).  
After recalling some useful facts from the 
conifold geometry \cite{co}, we solve (\ref{finale}). 
The metric is further constraint by the Bianchi identity 
on the torsion
\be
dH=\rho_5\,,
\label{bianchiH}
\ee
where $\rho_5$ is the properly normalized density of the NS5 branes. 
In this paper we assume $\rho_5=0$, except for possible delta-function
sources.  

A singular conifold can be described by a quadric in $\C_4$
\be
X Y-U V=0\,.
\label{con}
\ee 
A small resolution of the cone 
is obtained by replacing the 
equation (\ref{con}) by the pair of equations
\be
\tK:\qquad W\pmatrix{\l_1\cr\l_2}=0\,, \qquad  
\pmatrix{\l_1\cr\l_2}\in \P_1\,,
\label{recon}
\ee
where the matrix 
\be
W=\pmatrix{X&U\cr V&Y}\,,
\label{wdef}
\ee
has rank $1$, except when all of $X,Y,U,V$ vanish where it has rank $0$.
Away from the apex of the cone (\ref{recon}) determines a unique 
point on  $\P_1$. At the apex of the cone 
(\ref{recon}) defines the entire $\P_1$. Let $H_+$and $H_{-}$ 
be two coordinate patches  covering $\P_1$ with local 
coordinates $\lambda\equiv {\l_2\over \l_1}$, 
$\l_1\ne 0$ and $\mu\equiv {\l_1\over \l_2}$,
$\l_2\ne 0$ respectively. On $H_+$, (\ref{recon}) implies 
\be
W=\pmatrix{-U\l&U\cr -Y\l&Y}\,,
\label{WHp}
\ee
so we can choose $(U,Y,\l)$ as our complex coordinates.
Similarly, on  $H_{-}$ we have 
\be
W=\pmatrix{X&-X\mu\cr V&-V\mu}\,,
\label{WHm}
\ee
so we can choose  as complex coordinates $(V,X,\mu)$. 
On $H_+\cap H_-$ the transition function is given by
\be
(V,X,\mu)=(-Y\l,-U\l,1/\l)\,.
\label{transition}
\ee
In above coordinates the 
holomorphic $(3,0)$ form takes a simple form
\be
\w=dU\wedge dY\wedge d\l=dV\wedge dX\wedge d\mu\,.
\label{defw}
\ee 
Resolved conifold has global $SU(2)\times SU(2)$ symmetry with the 
following action 
\bea
W&\to& L W R^+\,,\cr\cr
\pmatrix{\l_1\cr \l_2}&\to& R\pmatrix{\l_1\cr \l_2}\,,
\label{actionSU2}
\eea
where $L,R$ are independent $SU(2)$ matrices. The three form $\w$
of (\ref{defw}) is invariant under the symmetries of the resolved 
conifold, and up to a c-number factor is unique\footnote{Any other 
holomorphic three form $\tilde{\omega}$ would differ from (\ref{defw}) 
by a  holomorphic
$SU(2)\times SU(2)$ invariant function $f(W):\ \tilde{\omega}=
f(W) \omega$, where $W$ is given by (\ref{WHp}), (\ref{WHm}) on $H_+,\ H_-$
correspondingly. With the transformation law (\ref{actionSU2}), 
$f$ can depend only on $\det[W]$: $f(W)\equiv 
\tilde{f}(\det[W])=\tilde{f}(0)$, which is constant.}.

The most general $SU(2)\times SU(2)$ invariant Hermitian metric 
on $\tK$ takes the form
\be
ds_6^2=f_1\ \tr\left(dW^+dW\right)+f_2\ |\tr\left(W^+dW\right)|^2+
f_3\ {|d\l|^2\over (1+|\l|^2)^2}\,,
\label{metricKt}
\ee
where $f_i=f_i(x)$ are scalar functions of 
\be
x\equiv \tr\left(W^+ W\right)\,. 
\label{fi}
\ee
Consider first condition coming from the gravitino variation 
(\ref{finale}). The norm of the holomorphic three form (\ref{defw}) in 
(\ref{metricKt}) evaluates to 
\be
||\w||=\sqrt{6}\ {\rm det}^{-1/2}(g_{a\bar{b}})=\sqrt{6}\
\biggl[f_1(f_1+f_2 x)(f_1 x+f_3)\biggr]^{-1/2}\,.
\label{normw}
\ee
As a shorthand notation we define
\be
g\equiv{\rm det}(g_{a\bar{b}})=f_1(f_1+f_2 x)(f_1 x+f_3)\,.
\label{detg}
\ee
Now, 
\be
i\left(\bar{\partial}-\partial\right)\ln||\w||=
-{i\over 2}\ [\ln g]' \left[\tr\left(dW^+ W\right)-
\tr\left(W^+ dW\right)\right]\,,
\label{derw}
\ee
where the prime denotes derivative with respect to $x$.
After some straightforward, though rather tedious algebra 
we find 
\be
d^+J=i\ {2 f_1 f_1' x+f_3 f_1'+f_1 f_3'-f_2 f_3-2 f_1 f_2 x\over 
f_1(f_1 x+f_3)}\ \left[\tr\left(dW^+ W\right)-
\tr\left(W^+ dW\right)\right]\,.
\label{d+J}
\ee 
With (\ref{derw}) and (\ref{d+J}), eq.~(\ref{finale}) gives 
\be
\biggl[f_1(f_1+f_2 x)(f_1 x+f_3)\biggr]'+2(f_1+x f_2)
\biggl[(f_2-f_1')(f_3+2f_1 x)-f_1 f_3'\biggr]=0\,.
\label{eq1}
\ee 
Once the metric (\ref{metricKt}) is specified, 
we can determine torsion following (\ref{h}). 
We will not give the complete expression here, just 
mention that the Bianchi identity on $H$ (in the absence
of NS5 branes) results in the following constraints
\bea
&&\biggl[(f_1'-f_2)x^2\biggr]'=0\,,\cr\cr
&&f_3'=2x(f_2-f_1')\,.
\label{eq2}
\eea
Finally, the dilaton is determined by (\ref{wfin})
\be
\phi={\rm const}+{1\over 4}\ln\biggl[f_1(f_1+f_2 x)(f_1 x+f_3)\biggr]\,.
\label{dilaton}
\ee
Eqs.~(\ref{eq1}) and (\ref{eq2}) represent complete set of 
constraints that determine supersymmetric compactifications 
on type IIB string theory on resolved conifolds with torsion.

To write down the metric (\ref{metricKt}) explicitly we will
parameterize $W$ in terms of two sets of Euler angles\footnote{
This parametrization first appeared in \cite{pzt}.}
\bea
U&=&r e^{i(\psi+\phi_1+\phi_2)/2}\cos{\theta_1\over 2}
\cos{\theta_2\over 2}\,,\cr
Y&=&r e^{i(\psi-\phi_1+\phi_2)/2}\sin{\theta_1\over 2}
\cos{\theta_2\over 2}\,,\cr
\l&=&e^{-i\phi_2}\tan{\theta_2\over 2}\,.
\label{param}
\eea
The metric on $\tK$ is then given by
\bea
ds_6^2&=&(dr)^2 \left(f_1+f_2 x\right)+{f_1 x\over 4}\left(
d\theta_1^2+\sin^2\theta_1\ d\phi_1^2\right)+{f_1 x+f_3\over 4}\left(
d\theta_2^2+\sin^2\theta_2\ d\phi_2^2\right)\cr
&&+{f_1 x+f_2 x^2\over 4}
\left(d\psi+\sum_{i=1}^2\ \cos\theta_i\ d\phi_i\right)^2\,,\qquad 
x=r^2\,.
\label{metricang}
\eea
From the Bianchi constraints (\ref{eq2}), away from the 5-brane 
sources, we find 
\be
\left[f_3' x\right]'=0\,,
\label{bcon}
\ee
with the most general solution 
\be
f_3=c_1\ln x+c_2\,,
\label{sol}
\ee
for some constants $c_i$.
Furthermore,
\be
f_2=f_1'+{f_3'\over 2x }\,.
\label{f2}
\ee
Given (\ref{sol}) and (\ref{f2}), (\ref{eq1}) gives an ordinary 
(nonlinear) differential equation on $f_1$. Some general conclusions
concerning supergravity backgrounds discussed here could be reached 
without explicitly solving the resulting equation. 
Most importantly,  with (\ref{sol}) we immediately see from 
(\ref{metricang}) that unless $c_1=0$ the background geometry 
always has a naked singularity: radius squared of one of the two
$S^2$s (parameterized by $(\theta_i,\phi_i)$ for $i=1$ or $i=2$)
will necessarily become negative  for  $r>r_s$ (or $r<r_s$ 
depending on the sign of $c_1$) with some $r_s$. 
From (\ref{dilaton}), we see that in the same region the dilaton 
would become complex valued. When $c_1=0$ and $c_2$ being arbitrary,
eq.~(\ref{f2}) requires $\tK$ to be K$\ddot{\rm a}$hler; we find in this 
case torsion to vanish identically. This is the main result of the 
section: we showed that there are no nonsingular supersymmetric 
backgrounds of type IIB supergravity on resolved conifolds 
with nontrivial torsion. 

We can also make a few comments about the nature of this naked 
singularity.  First of all, following the definition of \cite{kl95},
we can see that this is a repulson singularity.  That is, a massive 
particle coupled to the Einstein metric follows a radial trajectory 
given by
\be
\tau = \int dr |g_{rr}|^{1/2} g_{tt}^{1/2} \left[ E^2 - g_{tt} 
\right]^{-1/2}\,,
\label{repulse}
\ee
where $E$ is the energy per unit mass.
Because $g_{tt}= e^{-\phi/2}$ blows up at the singularity (recall 
(\ref{dilaton})), the particle 
will always bounce away from the singularity in finite proper time.
This also implies that our solutions violate the criterion of 
\cite{mn0007}, because $g_{tt}$ in the Einstein frame is unbounded at the
singularity.   This means they cannot accurately describe the IR dynamics of
a dual gauge theory.  This is not to say that our singularities cannot
be resolved (indeed the CV-MN solution is the resolution) but rather that 
there is no chirally symmetric phase to the dual gauge theory at 
extremality.

\section{Supergravity dual of YM theory with unbroken chiral symmetry}

In \cite{mn0008} Maldacena and Nunez described gravitational 
solutions corresponding to a large number of NS fivebranes wrapping 
a two sphere. They argued that these solutions describe 
pure $\N=1$ super YM in the IR. More specifically, they described 
two solutions: one having a smooth geometry and broken 
$U(1)_R$ chiral symmetry of the dual gauge theory 
in the IR\footnote{The complex geometry and the 
supersymmetry of the smooth solution has been discussed in details 
in \cite{pt0012}.}, and a solution with a naked singularity 
in the IR, dual to the gauge theory phase with the unbroken 
chiral symmetry.  The latter could be understood as a 
supersymmetric vacuum of type IIB string theory on the 
resolved conifold with torsion. Using the results of the previous 
section, we show here that this is indeed the case.

A simple solution of (\ref{eq1}), (\ref{eq2}) is 
\bea
f_3&=&-2a^2\ln {x\over r_*^2}\,,\cr
f_2&=&-{2a^2\over x^2}\ln {x\over r_*^2}\,,\cr
f_1&=&{a^2\over x} \left(1+2\ln {x\over r_*^2}\right)\,,
\label{specific}
\eea 
where $a,r_*$ are constants. 
The ten-dimensional metric in the string frame is given by 
\bea
ds^2_{str}&=&\eta_{\mu\nu}dx^{\mu} dx^{\nu}-
a^2\Biggl[{(dr)^2\over r^2}+{1+2\ln{r^2\over r_*^2}\over 4}
\left(d\theta_1^2+\sin^2\theta_1\ d\phi_1^2\right)\cr
&&+{1\over 4} \left(d\theta_2^2+\sin^2\theta_2\ d\phi_2^2\right)
+{1\over 4}\left(d\psi+\sum_{i=1}^2\ \cos\theta_i\ d\phi_i\right)^2\ 
\Biggr]\,.
\label{partsol}
\eea
The torsion is 
\be
H={a^2\over 4}\left[\left(d\psi+\sum_{i=1}^2\ \cos\theta_i\ d\phi_i
\right)\wedge\left(\sin\theta_1\ d\theta_1\wedge d\phi_1
-\sin\theta_2\ d\theta_2\wedge d\phi_2\right)\right]\,,
\label{hspes}
\ee
and the dilaton
\be
e^{\phi}={\rm const}\ \left({1+2\ln{r^2\over r_*^2}
\over r^4/a^4}\right)^{1/4}\,.
\label{dilspes}
\ee
Note that the naked singularity is at $r_s$ 
\be
r_s=r_* e^{-1/4}\,.
\label{rsing}
\ee

We would like to match (\ref{partsol})-(\ref{dilspes}) with the 
UV behavior of the smooth CV-MN solution. The smooth ten dimensional 
solution in \cite{mn0008} is given by 
\bea
ds^2_{str}&=&\eta_{\mu\nu}dx^{\mu} dx^{\nu}-
N\Biggl[d\rho^2+e^{2g(\rho)}\left(d\theta^2+\sin^2\theta\ d\varphi^2\right)
\cr
&&+{1\over 4}\sum_a(w^a-A^a)^2\Biggr]\,,\cr
\cr
e^{2\phi}&=&{\rm const}\ {2 e^{g(\rho)}\over \sinh 2\rho}\,,\cr
\cr
\cr
H&=&{N\over 4}\Biggl[-(w^1-A^1)\wedge(w^2-A^2)\wedge(w^3-A^3)+
\sum_a F^a\wedge (w^a-A^a)\Biggr]\,,\cr
&& \label{malnu}
\eea 
where 
\bea
A^1&=&a(\rho)\,,\qquad A^2=a(\rho)\sin\theta\,,\qquad A^3=\cos\theta\,,\cr
a(\rho)&=&{2\rho\over \sinh 2\rho}\,,\cr
e^{2g}&=&\rho\coth 2\rho-{\rho^2\over \sinh^2 2\rho}-{1\over 4}\,,\cr
w^1+i w^2&=&e^{-i\psi}\left(d\tilde{\theta}+i\sin\tilde{\theta}d\phi\right)\,,
\qquad w^3=d\psi+\cos\tilde{\theta}d\phi\,.
\label{mn1}
\eea
In the limit $\rho\to \infty$ we find that both 
backgrounds agree, provided we identify 
\bea
\rho&=&\ln r\,,\qquad {\rm as }\ r\to \infty\,,\cr
N&=&a^2\,.
\label{ident}
\eea

\section{Towards finite temperature resolution of the 
IR singularity}
In the previous section we showed that the supergravity solution 
described by Maldacena and Nunez \cite{mn0008}
with unbroken $U(1)_R$ symmetry is in the class of type 
IIB compactifications on resolved conifold with nonvanishing torsion. 
All such solutions have a naked singularity in the IR which is the 
reflection of the fact that the chiral symmetry in the dual Yang Mills 
{\it must} be broken at low energies. In fact, the smooth 
CV-MN solution has this $U(1)_R$ symmetry broken to a $Z_2$, as 
predicted by the dual gauge theory. 

Rather similarly, 
the naked singularity of the KT geometry is resolved in 
\cite{ks0007}. A different mechanism for resolving this  singularity 
was proposed  in \cite{b0011}. As we expect that the chiral 
symmetry of the gauge theory is restored at a finite temperature 
$T_c$, we expect that there should exist a non-extremal generalization 
of the KT geometry with the regular horizon cloaking the singularity. 
Such regular Schwarzschild horizon should appear only for some 
finite Hawking temperature. This is a rather unusual  phenomenon
from the supergravity point of view. 
The non-BPS generalization presented in \cite{b0011} does not 
realize this proposal. As shown in \cite{new}, the horizon of the 
solution discussed there is singular for arbitrary non-extremality 
parameter. The horizon singularity can be traced back \cite{new} to the 
too restrictive requirement of the self-duality of the three form fluxes
off the extremality. On a more technical level, this corresponds to the 
fact that the $U(1)$ fiber of compact $T^{1,1}$ in the KT geometry 
was not ``squashed'' relative to the 2-spheres of $T^{1,1}$.
Such squashing does not violate  $U(1)_R$ chiral symmetry and is 
necessary for a non-BPS solution to have a nonsingular horizon. 
This has been  shown in \cite{shkt}. There, the daunting task of solving 
a coupled system of the second order differential equations 
describing regular non-extremal generalization of the KT 
geometry was approached with a beautiful physical insight:
in the KT-KS model the number of fractional D3 branes is fixed, while the 
number or regular ones changes logarithmically with the energy scale 
(radial coordinate); thus, if at the horizon of the non-BPS 
KT geometry the number of regular D3 branes is still large, one could imagine 
developing a perturbation theory around standard black D3 branes \cite{sh},
with the small parameter being the ratio of fractional and the regular 
D3 branes. Computation to the first order in this perturbation theory 
demonstrated the  chiral symmetry restoration
in the gravitational dual of the cascading gauge theory at high temperature,
with horizon  ``cloaking'' the naked singularity of the extremal KT solution
\cite{shkt}.

In this section we  construct 
nonsingular, non-BPS generalizations of the geometry 
(\ref{partsol})-(\ref{dilspes}). In the case of $T^{1,1}$
of the KT geometry, squashing  the $U(1)$ fiber off the extremality 
induced a source for the dilaton \cite{new,shkt}. That is, the constant
dilaton at $T=0$, should run for $T\ge T_c$. In our 
case, the dilaton is nontrivial even at the extremality. 
In constructing the appropriate non-BPS 
solution we assume that the $U(1)$ fiber (parameterized by 
$\psi$ in (\ref{partsol})) is not squashed relative to the 
$(\theta_2,\phi_2)$ sphere, as it is at the extremality.
This restriction substantially simplifies 
type IIB equations of motion, but unlike 
a  somewhat similar ansatz in \cite{b0011},   
leads to geometries with regular horizons.

In what follows we discuss the following non-extremal
generalization of (\ref{partsol})-(\ref{dilspes}):
\bea
ds^2_{E}&=&c_1(r)^2\left[\t_1^2 dt^2-d\bar{x}^2\right]
-c_1(r)^2 a^2\Biggl[{dr^2\over \t_2^2 r^2}+{h(r)\over 4}\left(
d\theta_1^2+\sin^2\theta_1\ d\phi^2_1\right)\cr
&&+{1\over 4}\left(
d\theta_2^2+\sin^2\theta_2\ d\phi^2_2\right)+{1\over 4}
\left(d\psi+\sum_{i=1}^2\ \cos\theta_i\ d\phi_i\right)^2\
\Biggr]\,,\cr
\cr
H&=&{a^2\over 4}\left[\left(d\psi+\sum_{i=1}^2\ \cos\theta_i\ d\phi_i
\right)\wedge\left(\sin\theta_1\ d\theta_1\wedge d\phi_1
-\sin\theta_2\ d\theta_2\wedge d\phi_2\right)\right]\,,\cr
\cr
\phi(r)&=&\phi(r)\,,
\label{nonextr}
\eea 
where the metric is given in Einstein frame, $\phi$ in the last 
equation denotes the dilaton.  Note that we used 
the same torsion as in the extremal case (\ref{hspes}).

Checking type IIB equations of motion is tedious, so we only 
outline the steps. We perform the computations in the orthonormal
frame with 
\bea
&&e^1=c_1\t_1 d t\,,\qquad e^{2\cdots 4}=c_1 d x^{1\cdots 3}\,,
\qquad e^5=c_1 a 
\ {dr\over \t_2 r}\,,\cr
&&e^6={c_1 a\over 2}\ \left(d\psi+\sum_{i=1}^2\ \cos\theta_i\ 
d\phi_i\right)\,,\qquad e^7={c_1 a h^{1/2}\over 2}\ d\theta_1\,,\cr
&&e^8={c_1 a h^{1/2}\over 2}\ \sin\theta_1\ d\phi_1\,,\qquad 
e^9={c_1 a\over 2}\ d\theta_2\,,\cr  
&&e^{10}={c_1 a\over 2}\ \sin\theta_2\ d\phi_2\,.
\label{framedef}
\eea 
With ansatz (\ref{nonextr}), the 3-form Maxwell equations and its 
Bianchi identity are satisfied automatically. The energy-momentum
tensor of the three form satisfies 
\be
T^{(3)}_{11}+T^{(3)}_{22}=0\,, 
\label{eqq1}
\ee
which requires the sum of the corresponding components of the 
Ricci tensor to vanish. This gives 
\be
\t_1' \t_2={A\over c_1(r)^8 h(r) r}\,,
\label{eqq2}
\ee
where (a constant)   $A$ is the non-extremality parameter. 
Turns out, all the other Einstein equations, and the 
dilaton equation, could be reduced to the following 
second order differential equations
\bea
0&=&A^2 \biggl[\ln F(y)\biggr]''-4 F^2(y)\left(h^2(y)+1\right)\,,\cr
\cr
0&=&A^2 \biggl[\ln h(y)\biggr]''-8 F^2(y)\left(h(y)-1\right)\,,\cr
\cr
0&=&A^2 \biggl(\left[F^2(y)\right]'\left[h^2(y)\right]' 
+F^2(y) \left(h(y)'\right)^2+2 h^2(y)\left(F(y)'\right)^2\cr
&&-2 F^2(y) h^2(y)\biggr)-8 F^4(y) h^2(y)\left(h^2(y)+2h(y)-1\right)\,,
\label{three}
\eea
where the derivatives are with respect to\footnote{Note that 
the ``good'' coordinate of \cite{new,shkt} is also proportional to
$\ln \t_1$. } 
\be
y\equiv \ln \t_1(r)\,,
\label{defy}
\ee
also
\bea
F(y(r))&=&c_1^8(r) \t_1(r)\,,\qquad h(r)=h(y(r))\,,\cr\cr
e^{-\phi/2}&=&c_1^2(r)\,.
\label{otherdef}
\eea
Note that dilaton is related to the warp factor of three space dimensions 
$\bar{x}$ as in the extremal case.

The system of differential equations (\ref{three}) is overdetermined. 
It is straightforward to check that it is actually compatible. 
Further assuming $F(y)=F(h(y))$ one can obtain from (\ref{three}) 
a second order nonlinear differential equation for 
$F(h)$
\bea
&&F'' F h^2\left[A^2+4 F^2\left(h^2+2h-1\right)\right]+F' h\biggl[
-F\left(4F^2\left(h^2-3h+4\right)-A^2\right)\cr
&&-F' h\left(8F^2\left(h^2-h+2\right)+A^2\right)
+8F'^2 F h^2(h-1)\biggr]\cr
&&-2F^4(h^2+1)\ =\ 0\,,
\label{fhe}
\eea
where the derivatives are with respect to $h$.
Above equation is solved with 
\be
F(h)=C\ h^{-1/2} e^{h/2}\,,
\label{fh}
\ee
for zero non-extremality parameter $A=0$, thus reproducing 
(\ref{partsol})-(\ref{dilspes}) directly from the type IIB 
equations of motion.  The constant $C$ in (\ref{fh}) depends 
on parameter $r_\star$ in (\ref{partsol}) and the bare string 
coupling $e^{\phi_0}$, $C=r_\star^2 e^{-2\phi_0}$.

In the remaining of this section we argue that 
our solution (\ref{three}) have a regular
horizon. It is easy to see that there will be a regular horizon 
at $r=r_h$, $\t_1(r_h)=0$, when $F(y(r))e^{-y(r)}$ and $h(r)$
are nonzero at  $r=r_h$. Really, in this case in the vicinity 
of $r_h$, we can introduce a well-defined coordinate $\eta\equiv 
\t_1$, so that the metric (\ref{nonextr}) can be written as 
\bea
ds^2_{E}&\approx&c_1(r_h)^2\left[\eta^2 dt^2-d\bar{x}^2\right]
-{c_1^{18}(r_h) a^2 h^2(r_h)\over A^2}\ d\eta^2\cr
&&-c_1(r_h)^2 a^2\Biggl[{h(r_h)\over 4}\left(
d\theta_1^2+\sin^2\theta_1\ d\phi^2_1\right)\cr
&&+{1\over 4}\left(
d\theta_2^2+\sin^2\theta_2\ d\phi^2_2\right)+{1\over 4}
\left(d\psi+\sum_{i=1}^2\ \cos\theta_i\ d\phi_i\right)^2\
\Biggr]\,,
\label{etametric}
\eea  
which clearly describes a nonsingular horizon at $r_h$, 
provided $c_1(r_h)$ and $h(r_h)$ are nonzero. 
From (\ref{defy}), as $r\to r_h$, $y\to -\infty$. 
So the existence of regular horizon implies  
\bea
&&F(y)\to \alpha_1 e^y\,,\cr
&&h(y)\to \alpha_2\,,\qquad {\rm as}\ y\to -\infty\,,  
\label{fhasympt}
\eea 
for some  positive constants $\alpha_i$. This boundary conditions 
are  compatible with  (\ref{three}), and allow to construct a
power series solution\footnote{The form of the series expansion is simplest 
to deduce by rewriting (\ref{three}) in terms of a new variable 
$x\equiv \alpha_1 e^y$.}
\bea
F(y)&=&\alpha_1 e^y\left[1+\sum_{k=1}^\infty q_k e^{2 k y}\right]\,,\cr
h(y)&=&\alpha_2\left[1+\sum_{k=1}^\infty p_k e^{2 k y}\right],\qquad (-y)
\gg 1\,,
\label{horass}
\eea
where the  first couple terms are given by
\bea
&&q_1=\left({\alpha_1\over A}\right)^2 \left(1+\alpha_2^2\right)\,,\qquad 
q_2=\left({\alpha_1\over A}\right)^4
\left(1+\alpha_2^2+\alpha_2^3+\alpha_2^4\right)\,,\cr
&&p_1=2\left({\alpha_1\over A}\right)^2 \left(\alpha_2-1\right)\,,\qquad 
p_2=\left({\alpha_1\over A}\right)^4 \left(\alpha_2-1\right)
\left(\alpha_2^2+3\alpha_2-1\right)\,.\cr
&&
\label{pq}
\eea 
Compatibility of the boundary conditions at the regular horizon 
with the equations of motion is a strong hint that such regular 
horizon indeed exists. This should be contrasted with the 
non-extremal generalization of the KT geometry discussed in 
\cite{b0011}. It is possible to show that boundary conditions 
at a regular horizon of the non-BPS generalization of the 
KT geometry with constant dilaton are actually incompatible with the 
equations  of motion\footnote{On one hand, this follows indirectly  
from the conclusion of \cite{new} that non-extremal deformations 
with constant dilaton produce {\it singular} horizons. One could also 
see this directly by repeating analysis discussed here.}. 

Once convinced that it is consistent to impose  boundary conditions 
of the regular horizon on (\ref{three}), the next step is 
to identify restrictions on $(\alpha_1,\alpha_2)$ coming from the 
asymptotic in the UV, where we expect to recover the extremal solution 
(\ref{fh}). The statement that parameters  $(\alpha_1,\alpha_2)$ 
compatible with the 
UV asymptotic exist at all,  is highly nontrivial. We show however, 
that it is actually true. 

From the first two equations in 
 (\ref{three}) and the boundary condition (\ref{horass}), it follows 
that both $F(y)$ and $h(y)$ are monotonic functions of $y$.
While $F(y)$ always increases, $h(y)$ increases for $\alpha_2>1$
and decreases for $\alpha_2<1$. When $\alpha_2=1$, $h(y)=1$ 
identically. In the UV, at the extremality, we have both 
$h$ and $F$ increasing, thus only $\alpha_2>1$ can be compatible with the 
UV asymptotic. In what follows we assume this is the case.
We would like to show now that both $F(y)$ and 
$h(y)$ become infinitely large at finite $y\equiv y_{UV}$.
This follows from the fact that as $r\to\infty$, $\t_1(r)$ approaches 
a constant --- actually one, for proper normalization of the 
non-BPS deformation. Really, as $h(y)\gg 1$, the first equation
in (\ref{three}) is well approximated by 
\be
A^2\biggl[\ln\biggl(F(y) h(y)\biggr)\biggr]''\approx 
4\left(F(y)h(y)\right)^2\,,
\label{assFh}
\ee
which can be  solved exactly
\be
F(y) h(y)\approx {A C_1  }\ {e^{y C_1+C_2}\over 1-e^{2(y C_1+ C_2)}}\,,
\label{Fhy}
\ee  
where $C_1$ and $C_2$  are integration constants. $C_1$ is positive 
as $F(y) h(y)$  is always positive, thus  $F(y) h(y)$ blows up at finite 
$y_{UV}=-C_2/C_1$. It is straightforward to check, that 
given (\ref{Fhy}), from the second equation in (\ref{three}),
\be
h(y)\approx -2\log\biggl[1-e^{2(y C_1+ C_2)}\biggr]\,,\qquad {\rm as}\ 
y\to y_{UV}-0\,.
\label{hy}
\ee
We already mentioned that proper normalization of the 
non-extremal deformation requires $y_{UV}\equiv \log[\t_1(r\to\infty)]=0$,
while we are finding that $y_{UV}=-C_2/C_1$. This is not a contradiction.
Note that equations (\ref{three}) are invariant under an arbitrary finite shift
of $y$. Shifts in $y$ can be absorbed into the multiplicative normalization 
of $\alpha_1$. Thus $y_{UV}=0$ uniquely fixes $\alpha_1$, and the only
free parameter is $\alpha_2$ constraint by  $\alpha_2>1$. 
We would like to show now that for any $\alpha_2>1$ we have asymptotically
(\ref{fh}) for specific $C$ determined by $\alpha_2$. 
This is easiest to see from equation (\ref{fhe}). We are looking for the 
most general solutions of (\ref{fhe}), such that 
\be
F(h)\to \infty\,,\qquad  {\rm as}\ h\to \infty\,.
\label{uvass}
\ee
Let's assume first that $A=0$. With (\ref{uvass}),  solution of 
(\ref{fhe}) is a power series in $e^{-h/2}$
\be
F(h)=d_0 e^{h/2} h^{-1/2}\biggl[1+d_1 e^{-h/2} h^{3/2}\biggl(1+
O\left(1/h\right)\biggr)+O\biggl(e^{-h}\biggr)\biggr]\,,
\label{fhass}
\ee      
where $d_0$ and $d_1$ are arbitrary integration constants. 
It is easy to see that (\ref{fhass}) also solves (\ref{uvass}) with 
$A\ne 0$ to the specified order, and the corrections to (\ref{fhe}) 
from finite $A$, show up as $\delta F(h)\sim A^2 e^{-h/2}$ corrections 
to (\ref{fhass}). Since the latter corrections are subdominant, 
they would generically fix $d_1$. This is indeed what we find
\be
F(h)=d_0 e^{h/2} h^{-1/2}\biggl[1-{A^2\over 4 d_0^2} e^{-h} h^{-2}\biggl(1+
O\left(1/h\right)\biggr)+O\biggl(e^{-2 h}\biggr)\biggr]\,.
\label{fhA}
\ee
That is, $d_1$ in (\ref{fhass}) is fixed to be zero for $A\ne 0$.  
Eq.~(\ref{fhA}) reproduces the extremal solution in the UV 
if we identify $C$ of (\ref{fh}) with $d_0$.  
Clearly $d_0$  depends on $(\alpha_2,A)$. The $A$ dependence 
is trivial. Note that the $A$ dependence of (\ref{three}) 
drops out if we redefine $F(y)\to F(y) A$. Thus 
\be
C\equiv d_0=A\ {\cal F}(\alpha_2)\,,
\label{C}
\ee
where ${\cal F}(\alpha_2)$ is some specific function. 
Unfortunately, we do not know the relevant exact analytical
solution of (\ref{three}), and thus we can not determine 
this function explicitly\footnote{Unlike solutions discussed 
in \cite{shkt} there is no small parameter here which could 
be used to set up a perturbation theory.}. 
Recall that for the extremal solution $C=r_*^2 e^{-2\phi_0}$, so 
from (\ref{C}) we identify 
\be
\alpha_2={\cal F}^{-1}\left({r_*^2\over A}e^{-2\phi_0}\right)\,,
\label{alF}
\ee
where we used ${\cal F}^{-1}$ to denote an inverse function to ${\cal F}$.
We argued previously that to have a regular horizon with proper UV asymptotic 
$\alpha_2>1$, thus
\be
{\cal F}^{-1}\left({r_*^2\over A}e^{-2\phi_0}\right)>1\,,
\label{fcc}
\ee
indirectly determines relation between the non-extremality 
parameter $A$ and the scale of the chiral symmetry breaking of
the extremal solution $\sim r_*$. 

Additionally, numerical results support our belief that solutions exist
which properly interpolate between the boundary conditions (\ref{horass}),
(\ref{pq}) at the
horizon and the appropriate UV asymptotic (\ref{fh}).  To do so, we set
initial conditions for $F$ and $h$ using the $p_1,q_1$ terms in the series
(\ref{horass}) at a value of $x=e^y$ such that the correction was small.  
Then we solved the first two equations of (\ref{three}) numerically (using
the variable $x$).  For all the cases studied, both $F\to \infty$ and 
$h\to \infty$ at a finite value of $x$, which can be shifted to $x=1$ by
the multiplicative normalization of $\alpha_1$ discussed above.  It is then
possible to check the UV asymptotic by making a log-log plot of
$F e^{-h/2}$ versus $h$ near $x=1$ and find the power law slope.  Again,
in all cases studied, we found a power law between $h^{-0.49}$ and 
$h^{-1/2}$, correctly reproducing the UV asymptotic (\ref{fh}).

To summarize, 
in this section we constructed a family of non-extremal deformations 
of the singular CV-MN solution. We argued that our deformations 
have a regular Schwarzschild horizon, and showed that given the 
scale of a naked singularity of the chirally symmetric CV-MN solution,
$r_*$, its non-singular non-extremal deformation exists only for a range 
of non-extremality parameter $A$ determined from (\ref{fcc}).

\section{Discussion}
In this paper we considered supersymmetric compactifications 
of type IIB string theory on resolved conifolds with torsion. 
We extended the no-go theorem stated previously for compact 
Hermitian six dimensional manifolds in \cite{compact} to the 
non-compact manifolds with the complex structure of the resolved
conifold, and showed that nonvanishing torsion leads to the 
naked singularities in the geometry. 
In classifying supersymmetric compactifications 
we have not found the nonsingular Chamseddine-Volkov-Maldacena-Nunez 
solution \cite{chv1,chv2,mn0008},
which was argued in \cite{pt0012} to be in the class  
of string backgrounds on Hermitian manifolds with torsion. 
As the infrared (small $r$) geometry of \cite{chv1,chv2,mn0008} 
is that of the deformed conifold, and thus its complex 
structure is  different from the complex structure  
of the resolved conifold, the latter fact is not a contradiction. 
Classification of supersymmetric compactifications on deformed conifolds 
with torsion should recover solution of \cite{chv1,chv2,mn0008} and might 
uncover other interesting nonsingular backgrounds. We plan to return 
to this problem in the future.

We discussed non-extremal generalizations of the CV-MN
solution describing the  unbroken symmetry phase  
of the dual Yang Mills theory. We argued (though not proven rigorously) 
that the simplest solution we found 
have regular horizon which develops only above some critical 
non-extremality. This is in accord with the gauge theory where 
the chiral symmetry restoration occurs at finite temperature. 
There are lots of open questions. First of all, it would be extremely 
interesting to determine the exact analytical form of the function 
$\cal F$ in  (\ref{C}), and thus {\it explicitly} prove the restriction on 
the non-extremality coming from the boundary conditions of the regular 
horizon. Maybe it is possible to solve (\ref{three}) 
exactly\footnote{It is possible to
reduce the first two second order equations in (\ref{three}) to first order 
equations; the third equation becomes then algebraic.}  
? We did not discuss the thermodynamical
quantities of the background. It would also be interesting to 
study the critical black hole in this non-extremal geometry, 
corresponding to the  gauge theory at the phase transition.  
Finally, the non-extremal generalization  we found is rather
special. It is thus interesting to find out how generic it is, and 
if other solutions exist, what is their interpretation.

\section*{Acknowledgments}
We wish to thank Jerome Gauntlett, Steven Gubser, Sunny Itzhaki, 
Igor Klebanov, Andrei Mikhailov, Amanda Peet and 
Joe Polchinski for valuable discussions. The work of AB was supported in 
part by NSF grants PHY97-22022 and PHY99-07949. The work of AF was 
supported by a National Science Foundation Graduate Research Fellowship.

\section*{Appendix}
We follow notations of \cite{schwarz83}. The metric signature is mostly 
minus and the Clifford algebra is 
\be
\{\hg^{r_i},\hg^{r_j}\}=2\eta^{r_i r_j}\,,
\label{g10}
\ee
where $r_i$ is the tangent space index and $\eta^{r_i r_j}$ is the 
Minkowski metric. We define
\bea
&&\hg^{r_1\cdots r_k}=\hg^{[r1}\hg^{r_2}\cdots\hg^{r_k]}\,,\cr
&&\hg^{11}=\hg^1\cdots \hg^{10}\,,
\label{multig}
\eea
where symmetrization of indices
is carried out with weight one: $[ab]={1\over 2!}(ab-ba)$.
The notation $\hg$ indicates a ten-dimensional 
quantity, while $g$ indicates  either  four or six
dimensional quantity. In a Majorana representation $\hg^1$ is antisymmetric 
and imaginary, and $\hg^2$ to $\hg^{10}$ are symmetric and imaginary. 
When  $r$ indices are used, $\hg$ matrices are purely numerical
(independent of the coordinates). When they are converted to greek indices 
with the 10-bein $\he^r_M$ or its inverse $\he^M_r$, they become 
field dependent functions.  
We use roman subindices 
$M,N,...=1,...,10$; $\mu,\nu=1,2,3,4$ and $m,n,...=5,...,10$.

We consider warped product geometries of the form 
\be
d\hs_{10}^2=e^{2A(y)}\eta_{\mu\nu}dx^{\mu}dx^{\nu}-e^{2 B(y)}
g_{mn}dy^mdy^n\,,
\label{am}
\ee
where $A(y)$, are $B(y)$ are warp factors that depend only 
on 6D indices. The minus sign in $(\ref{am})$ allows us to have 
six-dimensional metric of positive signature. The natural relation 
between 10- and 4- (6-) beins is 
\be
\he_{\mu}^r=e^A e_{\mu}^r\,,\qquad  \he_{m}^r=i e^B e_{m}^r\,,
\label{frames}
\ee  
where $i$ in (\ref{frames}) accounts for the change of signature 
$\hat{g}_{mn}=-e^{2B}g_{mn}$. From (\ref{frames}),
\be
\hg_m=i e^B \g_m\,,\qquad \hg^m=-i e^{-B} \g^m\,,
\label{g106}
\ee 
with 
\be
\{\g_m,\g_n\}=2g_{mn}\,.
\label{6antic}
\ee
Note that the $\g_m$ matrices are symmetric and real. 

For the $k$-forms we use notation 
$F_{(k)}={1\over k!}F_{M_1\cdots M_k} dx^{M_1}\wedge\cdots \wedge dx^{M_k}$.
We define 
\be
\slashed {\hat{F}}_{(k)}={1\over k!} F_{M_1\cdots M_k} \hg^{M_1\cdots M_k}\,.
\label{slash}
\ee
Note that if the $k$-form is nonzero only in six dimensions,
\be
\slashed {\hat{F}}_{(k)}=(-i)^k e^{-k B}\slashed {{F}}_{(k)}\,.
\label{s106}
\ee
Hodge duals are defined as 
\be
(\hat{\star}F)_{M_{k+1}\dots M_{10}}
={\sqrt{|\hat{g}|}\over k! } 
\hat{\e}_{M_{k+1}\dots M_{10}}{}^{M_1\cdots M_k}F_{M_1\cdots M_k}\,,
\label{hodge}
\ee   
and similarly for the six-dimensional Hodge dual $\star$.  
We take convention $\hat{\e}_{12\cdots 10}=+1$; also $\e_{5\cdots 10}=+1$. 
We  also need the adjoint exterior derivative operator, which in 
six-dimensions is defined as 
\be
d^+\equiv \star d\star\,.
\label{ds}
\ee

\end{document}